\documentclass{svproc}
\usepackage{url}

\usepackage{graphicx, multicol, footmisc}
\usepackage{lineno}
\usepackage{siunitx}
\usepackage{amssymb, mathtools}
\usepackage{amsmath}
\usepackage{todonotes}
\usepackage{booktabs}
\usepackage{xspace}
\usepackage[caption=false]{subfig}
\usepackage[english]{babel}
\usepackage{microtype}
\usepackage{spverbatim}
\usepackage[short, nocomma]{optidef}
\usepackage[misc]{ifsym}
\usepackage[hidelinks]{hyperref}
\graphicspath{{figures/}}

\newcommand{\WSP}[1]{\mbox{WSP}(#1)\xspace}

\newcommand{\T}{\ensuremath{\mathsf{T}}}

\makeatletter
\renewcommand{\fnum@figure}{Figure \thefigure}
\makeatother

\begin{document}
\mainmatter              % start of a contribution
\title{PaMILO: A Solver for Multi-Objective Mixed Integer Linear Optimization and Beyond}
\titlerunning{PaMILO}  % abbreviated title (for running head)
\author{Fritz Bökler \and Levin Nemesch$^{(\href{mailto:lnemesch@uni-osnabrueck.de}{\text{\Letter}})}$ \and Mirko H. Wagner}
\authorrunning{F. Bökler, L. Nemesch, M. H. Wagner} % abbreviated author list (for running head)
% \tocauthor{Fritz Bökler \and Levin Nemesch \and Mirko H. Wagner}
%
\institute{Universität Osnabrück, Germany,\\
            Theoretical Computer Science: \url{https://tcs.uos.de/}\\
            \email{\{fboekler|lnemesch|mirwagner\}@uni-osnabrueck.de}}

\maketitle              % typeset the title of the contribution

\begin{abstract}
% The abstract should summarize the contents of the paper
% using at least 70 and at most 150 words. It will be set in 9-point
% font size and be inset 1.0 cm from the right and left margins.
% There will be two blank lines before and after the Abstract. \dots
In multi-objective optimization, several potentially conflicting objective functions need to be optimized.
Instead of one optimal solution, we look for the set of so called non-dominated solutions.

An important subset is the set of non-dominated extreme points.
Finding it is a computationally hard problem in general.
While solvers for similar problems exist, there are none known for multi-objective mixed integer linear programs (MOMILPs) or multi-objective mixed integer quadratically constrained quadratic programs (MOMIQCQPs).
We present PaMILO, the first solver for finding non-dominated extreme points of MOMILPs and MOMIQCQPs.
It can be found on github under \texttt{\href{https://github.com/FritzBo/PaMILO}{github.com/FritzBo/PaMILO}}.
PaMILO provides an easy-to-use interface and is implemented in C++17.
It solves occurring subproblems employing either CPLEX or Gurobi.

PaMILO adapts the Dual-Benson algorithm for multi-objective linear programming (MOLP).
As it was previously only defined for MOLPs, we describe how it can be adapted for MOMILPs, MOMIQCQPs and even more problem classes in the future.

% We would like to encourage you to list your keywords within
% the abstract section using the \keywords{...} command.
\end{abstract}

\section{Introduction}

Many optimization problems can be seen as \emph{multi-objective problems}, where several potentially conflicting objective functions need to be optimized. The solution of such a problem consists not of one single optimal solution value, but instead a set of so called \emph{Pareto-optimal} solution values. A solution is Pareto-optimal, if an improvement in one objective has to worsen at least one other objective. Subclasses of multi-objective problems include \emph{multi-objective linear programming} (MOLP), \emph{multi-objective integer linear programming} (MOILP), \emph{multi-objective mixed integer linear programming} (MOMILP), and \emph{multi-objective mixed integer quadratically constrained quadratic programming} (MOMIQCQP). In this paper, we are interested in the so-called extreme points of MOMILPs and MOMIQCQPs. As MOLPs and MOILPs are special cases of MOMILPs, this also covers finding extreme points for those.

Especially MOMILPs are relevant problems in practice. For example in~\cite{antunes}, a model for power generation expansion planning is formulated as a MOMILP. The objectives consider the economic cost of expanding the infrastructure, the environmental impact and the cost of resulting economic damage. In~\cite{portfolio}, a MOMILP is used in a completely different field to optimize financial portfolios. A set of portfolios is calculated from which an investor can choose which one fits best to their preferences.

Extreme points are an especially interesting subset of the Pareto-optimal solution: If a decision process mirrors reducing multiple criteria into one dimension through a concave function, an optimal solution can always be represented by exactly one specific extreme point~\cite{benson1995concave}. Searching for extreme points can be seen as equal to parametric optimization, extreme points are called \emph{break points} there.
There are no extreme point solvers for MOMILPs or MOMIQCQPs and only few for MOLPs or MOILPs: \emph{bensolve}~\cite{bensolve} and \emph{inner}~\cite{inner} are solvers for MOLPs, \emph{PolySCIP}~\cite{polyscip} is able to find extreme points of both MOLPs and MOILPs.

\noindent
\textbf{Contribution.} We introduce \emph{PaMILO} (Parametric Mixed Integer Linear Optimization), a new solver for MOMILPs and MOMIQCQPs. It can read most common input formats and provides an easy-to-use console interface for practitioners.
The only requirement is the existence of a tight lower bound for each objective, a so-called \emph{ideal point}.
The existence of it is an assumption also made by the aforementioned state-of-the-art solvers for MOLPs and MOILPs.
Which is reasonable, as practical instances usually have an ideal point.
The algorithmic approach of PaMILO is based on the Dual-Benson algorithm.
We explain, how the Dual-Benson algorithm is adapted for MOMILPs and MOMIQCQPs.
Furthermore, through this technique the Dual-Benson algorithm can be adapted to all kinds of multi-objective optimization problems.

\section{Definitions}\label{sec:defs}

In MOMIQCQP, problems are of the form:
\begin{mini*}
    {}{f_i(x) \coloneqq x^\T P_i x + c_i^\T x \hspace{20pt}i=1,\ldots, d}{}{}
    \addConstraint{x^\T Q_j x + a_j^\T x}{\leq b}{\hspace*{25pt}j=1,\ldots, m}
    \addConstraint{x\in \mathbb{R}^\ell} {\times \mathbb Z^{\ell'},}
\end{mini*}
\noindent where $d,\ell,\ell',m\in\mathbb{N}$ and $n=\ell+\ell'$.
 For an index $i=1,\ldots, d$, we denote by $P_i\in\mathbb{Q}^{n\times n}$ the \emph{quadratic objective function matrix} $i$, and by $c_i\in\mathbb{Q}^{n}$ the \emph{objective function vector} $i$. For an index $j=1,\ldots,m$, we denote by  $Q_j\in\mathbb{Q}^{n\times n}$ the \emph{quadratic constraint matrix} $j$, and by $a_j\in\mathbb{Q}^{n}$ the \emph{constraint vector} $j$.
A single solution $x$ is \emph{feasible} if it fulfills all constraints, and we denote the set of all feasible solutions by $\mathcal{X}$.
The set of all feasible solutions in objective space $\{f(x) : x\in\mathcal{X}\}$ is called $\mathcal{Y}$.
For MOLP, MOILP and MOMILP all quadratic matrices $P_{i=1,\ldots,d}$ and $Q_{i=1,\ldots,m}$ are $\mathbf{0}$. The objective function then becomes $Cx$ with $c_{i=1,\ldots,d}^\T$ as the rows of $C$, and the constraints can be described through $Ax \leq b$ with $a_{j=1,\ldots,m}^\T$ as the rows of $A$.
Furthermore, in MOLP all variables are continuous and in MOILP all variables are integers.
Without loss of generality, we only consider minimization problems here.
But PaMILO is also able to solve problems with maximization objectives.

A point $y^*\in\mathcal{Y}$ is called \emph{non-dominated} if there is no $y\in\mathcal{Y}\backslash\{y^*\}$ with $y_i\leq y^*_i\ \forall i=1,\ldots,d$. In general, there is not a single non-dominated point, but a (not necessarily finite) set of non-dominated points. We call this the \emph{non-dominated set}, or $\mathcal{Y}_N$.

\begin{figure}[tb]
    \centering
    \subfloat{
        \centering
        \includegraphics[width=0.275\textwidth]{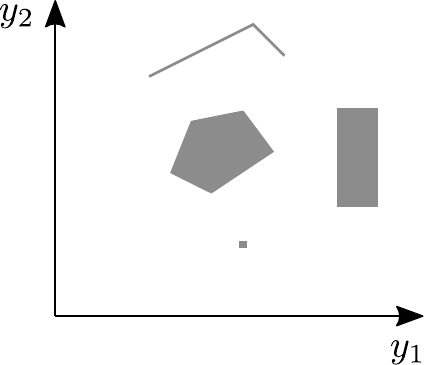}
    }
    \hspace{15pt}
    \subfloat{
        \centering
        \includegraphics[width=0.275\textwidth]{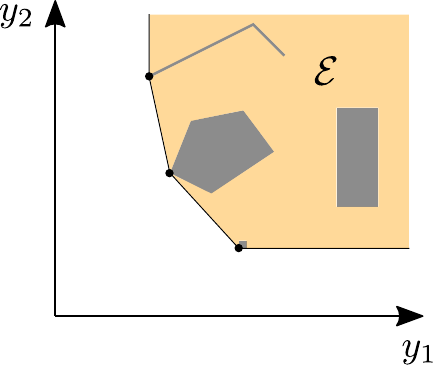}
    }
    \caption{Example of a MOMILP and its Edgeworth-Pareto hull.}\label{fig:ep_example}
\end{figure}

PaMILO searches for a subset of $\mathcal{Y}_N$, the non-dominated extreme points.
To properly define an extreme point, we first define the \emph{Edgeworth-Pareto hull} of a MOMIQCQP: $\mathcal{E}\coloneqq \text{conv}(\mathcal{Y})+\mathbb{R}_{\geq0}^d$.
Then, the \emph{non-dominated extreme points} are the vertices of $\mathcal{E}$.
A vertex of the (convex) set $\mathcal{E}$ is a point $y\in\mathcal{E}$ such that there is a valid hyperplane $H$ with $H\cap\mathcal{E}=\{y\}$.
PaMILO finds a finite representation of $\mathcal{E}$.
If the number of extreme points is finite itself, this representation is the exact set of extreme points.
In MOMILP, this set is always finite.
But there are MOMIQCQPs in which it may be infinite.
Figure~\ref{fig:ep_example} shows a MOMILP and its Edgeworth-Pareto hull.

A special case of the Edgeworth-Pareto hull is the \emph{upper image} for MOLPs. It is defined as $\mathcal{P}\coloneqq\mathcal{Y}+\mathbb{R}_{\geq0}^d$~\cite{heyde2008geometric}. In MOLP, $\mathcal{Y}$ is convex by itself and, thus, every point of $\mathcal{Y}_N$ lies on the boundary of $\mathcal{P}$. The same does not hold for MOMIQCQP, points of $\mathcal{Y}_N$ can lie inside $\mathcal{E}$.

The extreme points can also be characterized through the \emph{weighted-sum problem} (WSP) of a MOMILP.
Given a weight vector $w \in \mathcal{W}=\{w\in\mathbb{R}_{\geq0} : \sum_{i=1}^d w_i=1\}$, it is $\WSP{w}\coloneqq\min_{x\in\mathcal{X}}w^\T f(x)$.
An extreme point is a point $y\in \mathcal{Y}$ where a $w\in\mathcal{W}$ exists, so that the weighted sum in objective space is only minimal for this point.
Simply solving $\WSP{w}$\ might result in a dominated point in some special cases.
While PaMILO is able to handle these special cases, here we simply assume they do not appear.

\section{Algorithm}\label{sec:algorithm}

PaMILO adapts the Dual-Benson algorithm, which we briefly describe in this section. Originally, the Dual-Benson algorithm is only defined for MOLPs. In Subsection~\ref{sec:adapt} we explain how to also apply it to MOMILPs and MOMIQCQPs. A detailed description of the algorithm and its theoretical preliminaries can be found in \cite{ehrgott2012dual,hamel2014benson,heyde2008geometric}. The algorithm operates on the \emph{lower image} of the problem. The lower image $\mathcal{D}$ is defined in~\cite{heyde2008geometric} as
\[\mathcal{D}\coloneqq\left\{\left(w_1,\ldots,w_{d-1}, b^\T u\right) : w\in\mathcal{W}, A^\T u = C^\T w, u\in\mathbb{R}_{\geq 0}^m\right\}.\]

Heyde and Löhne showed in~\cite{heyde2008geometric} that the lower image is geometrically dual to the upper image. Particularly, every facet in the lower image corresponds to an extreme point in the upper image and vice versa. Hence, by enumerating the facets of the lower image, we obtain the extreme points of the MOLP.

The algorithm works by iteratively refining an outer approximation of $\mathcal{D}$. The initial outer approximation consists of one trivial facet of $\mathcal{D}$ and the boundaries given by $w\in\mathcal{W}$. The algorithm then improves this approximation by subsequently enumerating unvisited vertices of the approximation. A vertex is \emph{unvisited} if it was not previously used in an iteration. In each iteration, the algorithm picks an unvisited vertex $v$ of the approximation and shoots a ray down onto the boundary of $\mathcal{D}$. Shooting down a ray from $v$ is the same as solving $\WSP{v_1,\ldots,v_{d-1},1-\sum_i^{d-1}v_i}$. Either, the ray hits $\mathcal{D}$ directly at $v$ and confirms that $v$ is an extreme point of $\mathcal{D}$, or the ray hits a new point on the boundary of $\mathcal{D}$. If such a point is hit, a new facet supporting inequality of $\mathcal{D}$ is constructed that cuts off $v$. This improved approximation of $\mathcal{D}$ has new unvisited vertices for further iterations, so a vertex enumeration is done. When no unvisited vertex remains, the final approximation is the lower image itself. Because of geometric duality, this also gives us the vertices and facets of the upper image. Thus, we have found all non-dominated extreme points.

\begin{figure}[tb]
    \centering
    \subfloat[A ray is shot down from a dual vertex and finds a new dual facet.]{
        \centering
        \includegraphics[width=0.275\textwidth]{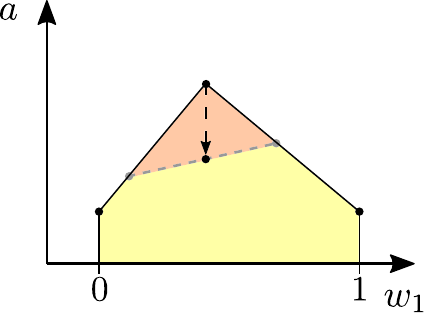}\label{fig:iter_demo1}
    }
    \hspace{15pt}
    \subfloat[The new facet is added to the dual approximation.]{
        \centering
        \includegraphics[width=0.275\textwidth]{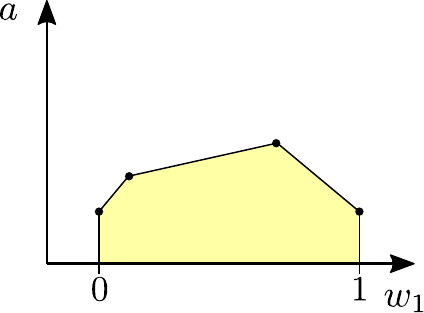}\label{fig:iter_demo2}
    }
    \caption{Example of one iteration of the Dual-Benson algorithm. The approximation (red) of $\mathcal{D}$ (yellow) becomes more accurate with the new facet.}\label{fig:iter_db}
\end{figure}

\subsection{Generalizing the Dual-Benson Algorithm}\label{sec:adapt}

As mentioned before, we can adapt the Dual-Benson algorithm to also find extreme points of MOMILPs and MOMIQCQPs. A previous generalization to multi-objective combinatorial problems was done by Bökler and Mutzel in~\cite{Fritz15}. Our adaption is based upon the ideas developed in~\cite{linzer2022}.

The Edgeworth-Pareto hull $\mathcal{E_I}$ of a MOMILP $\mathcal{I}$ (under assumption of an ideal point) is a convex polyhedron with a recession cone equal to the positive orthant.
The number of its vertices is finite.
Thus, there is a MOLP $\mathcal{J}$ so that its image $\mathcal{Y_J}$ is equal to $\mathcal{E_I}$.
$\mathcal{J}$ can be constructed by simply using the facets of $\mathcal{E_I}$ as constraints and making $C_{\mathcal{J}}$ the identity matrix.
This MOLP has an upper image $\mathcal{P_J}$ and a corresponding lower image $\mathcal{D_J}$.
As $\mathcal{P_J}$ has the same vertices as $\mathcal{E_I}$, a Dual-Benson algorithm operating on $\mathcal{J}$ finds the extreme points of $\mathcal{I}$.

There is a catch though, since we do not actually know $\mathcal{J}$.
But we do not need to:
The only time the Dual-Benson algorithm interacts with $\mathcal{J}$ is when shooting down rays onto the boundary of $\mathcal{D_J}$.
But ray shooting is the same as solving weighted-sum problems.
Thus, a weighted-sum oracle for $\mathcal{I}$ is sufficient to realize ray shooting onto $\mathcal{D_J}$.
With such a weighted-sum oracle, the Dual-Benson algorithm is able to find the extreme points of the MOMILP $\mathcal{I}$.

For MOMIQCQPs, the same arguments can directly be applied if the number of extreme points is finite.
But there also are MOMIQCQPs with an infinite number of extreme points.
In this case, the Dual-Benson algorithm would not terminate.
This can be avoided by only adding dual facets when the length of the ray hitting them is above a threshold value $\varepsilon>0$.
The resulting outer approximation of the lower image corresponds to an inner approximation of the actual Edgeworth-Pareto hull~\cite{ulus14}.
The extreme points of this inner approximation are a finite subset of the extreme points of $\mathcal{E}$.

As we never use distinct characteristics of MOMILPs or MOMIQCQPs, this argumentation can be generalized for all possible multi-objective problems.
As long as a weighted-sum is computable, a Dual-Benson algorithm using a weighted-sum oracle is able to find the extreme points or at least the extreme points of an approximation of $\mathcal{E}$.

\section{Implementation}\label{sec:implementation}

PaMILO is implemented in C++17.
Installation can be easily done by using cmake.
After installation, PAMILO provides an easy-to-use command-line interface.
Pamilo can be found on github under \texttt{\href{https://github.com/FritzBo/PaMILO}{github.com/FritzBo/PaMILO}}.
For academic purposes, it can be used subject to the MIT software license.

The weighted-sum problems are solved through CPLEX or Gurobi.
Due to the need to solve multi-objective problems, recent versions of the solvers are required.
For technical reasons, only the Gurobi version can handle MOMIQCQPs.

To minimize numerical error, instances undergo preprocessing.
If the value ranges of objectives in the non-dominated points differ strongly, the number of numerical errors increases.
PaMILOs preprocessing tries to normalize these ranges for each objective function.
Through this, we observe a significant decrease of numerical issues in practice.

Many formats are supported for input files, the most well known are the \texttt{.lp} and the \texttt{.mps} formats.
The output consists of three files, all beginning with a user defined output name as prefix.
The \texttt{*\_sol} file contains all extreme points and one solution corresponding to each in \texttt{.json} format, the \texttt{*\_log} file contains logging information of PaMILO, and the \texttt{*\_(cplex|gurobi)} file contains the logging output from the respective solver.

\section{Computational Results}

To get an impression of PaMILOs performance on established benchmark instances, we compared it to two other start-of-the-art solvers: bensolve~\cite{bensolve} and PolySCIP~\cite{polyscip}.
PolySCIP is used with CPLEX internally and PaMILO with Gurobi, the best variant for the respective solver.
We used the Dual-Benson algorithm as bensolves algorithm.
Computations were done on an Intel Xeon GOLD 6134 CPU with 256 GB RAM, and all instances had a time limit of 30 minutes.

Experiments for MOLPs use instances from~\cite{Fritz15}.
They show a very clear superiority of bensolve over PolySCIP and PaMILO on all instances.
For MOILPs, we generated instances according to the scheme for general MOILPs described in~\cite{kirlik2015computing}.
On those, PaMILO is often faster than PolySCIP while finding similar sets of extreme points.
For MOMILPs, we generated instances according to the scheme described in~\cite{mavrotas1998branch}.
The instances we generated were much bigger than the ones used in~\cite{mavrotas1998branch}, but PaMILO is still able to find extreme points.
A brief look on some of the computational results for MOILPs and MOMILPs is given in Table~1. %ugly, but who cares

In general, we observe that the most cost expensive part of the calculations are the vertex enumerations.
In theory, the computational effort to enumerate the vertices can grow exponentially in the number of objectives.
We observe such growth with PaMILO for many instances.
But for a fixed number of objectives, we also observe a delay between subsequent outputs that is only incremental polynomial in the oracle calls consistent with~\cite{bokler2018output}.

\setlength{\abovetopsep}{-4.69pt}
\begin{table}[!b]\label{table::experiments}
    \caption{Computational results for MOILPs and MOMILPs.
    Time and number of extreme points ($\mathcal{Y_\mathrm{Ex}}$) are median.
    Entries in $()$ are reduced data because of timeouts.}
    \begin{tabular*}{\textwidth}{@{\extracolsep{\fill}} l r r  r r r  r r r} \toprule
        \multicolumn{3}{c  }{Instances}  & \multicolumn{3}{c }{PaMILO} & \multicolumn{3}{c }{PolySCIP}\\ \cmidrule[0.55pt](){1-3}\cmidrule[0.55pt](){4-6}\cmidrule[0.55pt](){7-9}z
        type & $d$ & $n$ & time [s] & $|\mathcal{Y_\mathrm{Ex}}|$ & \# & time [s] & $|\mathcal{Y_\mathrm{Ex}}|$ & \#  \\ \midrule
         MOILP & $3$ & $100$ & $\mathbf{184.19}$ & $(28)$ & $(27/30)$ & $712.66$ & $(24)$ & $(22/30)$ \\
          & $4$ & $60$ & $\mathbf{191.8}$ & $(90.5)$ & $(18/30)$ & $828.4$ & $(67.5)$ & $(18/30)$ \\
          & $5$ & $50$ & $\mathbf{570.6}$ & $(139)$ & $(26/30)$ & --- & $(61)$ & $(13/30)$ \\ \hline
         MOMILP & $3$ & $300$ & $\mathbf{412.29}$ & $10074.5$ & $30$ & --- & --- & --- \\
          & $4$ & $200$ & $\mathbf{781.79}$ & $(24930.0)$ & $(29/30)$ & --- & --- & --- \\
          & $5$ & $40$ & $\mathbf{208.67}$ & $21546.0$ & $30$ & --- & --- & --- \\
        \bottomrule
    \end{tabular*}
\end{table}

To the best of our knowledge, there are no benchmark instance sets for MOMIQCQPs described in literature.
Our experiments aim at giving a first impression.
As there are now several extreme point solvers available, a comprehensive computational study on their capabilities is a goal for future research.

\section{Conclusion}

PaMILO is the first solver able to find the extreme points of MOMILPs and MOMIQCQPs and proves to be comparable to start-of-the-art solver PolySCIP for MOILPs.
Thus, it can be an important new addition to the repertoire of practitioners.

\bibliographystyle{spmpsci}
\bibliography{bib}

\end{document}